\documentclass[aps,prb,reprint,superscriptaddress]{revtex4-1}

\usepackage{lipsum}
\usepackage{graphicx}
\usepackage[english]{babel}  
\usepackage[utf8]{inputenc} 
\usepackage{amsmath,amssymb}
\usepackage{bm}
\usepackage[overload]{empheq}
\usepackage{subfigure}
\usepackage{numprint}
\usepackage{xcolor}
\usepackage{siunitx}

\usepackage{CJKutf8}

\newcommand{\expo}{\mathrm{e}}
\newcommand{\ic}{\rm i}

\newcommand*\conj[1]{
  \vbox{
  \hrule height 0.3pt
  \kern0.5ex
  \hbox{
  \kern-0.4em
  \ifmmode#1\else\ensuremath{#1}\fi
   \kern-0.em
  }
 } 
}

 \newcommand{\B}{\bm}

\newcommand\abs[1]{\left|#1\right|}

\begin{document}

\begin{CJK}{UTF8}{gbsn}

\title{A class of invisible inhomogeneous media and the control of electromagnetic waves}
\author{B. Vial}
\email{b.vial@qmul.ac.uk}
\author{Y. Liu (刘泱杰)}
\affiliation{School of Electronic Engineering and Computer Science, 
Queen Mary University of London, 
London E1 4NS, United Kingdom}

\author{S. A. R. Horsley}
\author{T. G. Philbin}
\affiliation{Department of Physics and Astronomy, 
University of Exeter, Stocker Road, Exeter, EX4~4QL, United Kingdom}

\author{Y. Hao}
\affiliation{School of Electronic Engineering and Computer Science, 
Queen Mary University of London, 
London E1 4NS, United Kingdom}

\date{\today}


\begin{abstract}
We propose a general method to arbitrarily manipulate an electromagnetic wave propagating in a two-dimensional medium, 
without introducing any scattering. This leads to a whole class of isotropic spatially varying permittivity and permeability profiles 
that are invisible while shaping the field magnitude and/or phase. In addition, we propose a metamaterial structure working in the infrared 
that demonstrates deep sub-wavelength control of the electric field amplitude and strong reduction of the scattering. This work 
offers an alternative strategy to achieve invisibility with isotropic materials and paves the way for tailoring the propagation of 
light at the nanoscale.
\end{abstract}

\maketitle

\end{CJK}

\section*{Introduction}

In recent years, the introduction of Transformation Optics has shed a new light on the propagation of electromagnetic 
waves in complex media and has proven to be an intuitive yet powerful tool for 
engineering the flow of light at the sub-wavelength scale \cite{Pendry1780,LeonhardtPhilbin2006,Leonhardt1777}. 
The theory is based on the invariance of Maxwell's equations under a change of coordinates, 
resulting in equivalent permittivity and permeability profiles that are generally anisotropic, spatially varying and sometimes singular. 
Perhaps the most popular application has been an invisibility cloak, which has been realized experimentally in various frequency regimes 
for two dimensional and three dimensional setups \cite{Schurig977,valentine2009optical,ergin2010three} 
thanks to the development of metamaterials and advanced manufacturing techniques \cite{chen2010transformation}. 
However, the complexity of the required material properties makes practical realisation a hard task, 
while the use of resonant meta-atoms to reach extreme parameters results usually in a narrow 
frequency band of operation \cite{Oscar2012,Oscar2013}. There is thus a critical need for other approaches to achieve invisibility at least to reduce 
diffraction significantly such as mantle cloaking \cite{AluMantle2009}, optimized dielectric covers \cite{Sigmund2011,Vial2015} or by 
introducing gain \cite{Lin2011,Mostafazadeh2013}. 
Quite paradoxically, although it is a very common phenomenon in wave physics, relatively little is known regarding what does or does not cause 
scattering when the material properties are allowed to vary rapidly in space \cite{Berry1990,Horsley2016,Philbin2016,horsley2015spatial}. 
Finally, there is an ever increasing demand for controlling optical fields 
at the nanoscale for applications ranging from medical diagnostics and sensing to optical devices and
optoelectronic circuitry \cite{zeng2014nanomaterials,Singh,Ozbay189,li2008harnessing}. In particular, local field enhancement is of paramount 
importance in phenomena such as surface enhanced Raman scattering (SERS) \cite{SERS2013,stiles2008surface}, improved non-linear effects 
\cite{EnhancNonlinear1,novotny2012,nphotonZayats}, optical antennae and the 
control of the local density of states \cite{hoppener2012self,belacel2013controlling}.\\
In this paper we present a general purpose method to control the amplitude and/or phase of a wave propagating 
in a two dimensional (2D) inhomogeneous isotropic medium. Although we focus our attention on media 
that does not scatter an incident plane wave while producing a 
specified amplitude and/or phase, the technique might be extended to arbitrary incident fields as well as to control the scattering pattern. 
In addition, the method is not based on the geometrical optics approximation and is valid at every frequency.\\

\section{Governing equations}
We consider here linear, isotropic, lossless and possibly dispersive materials characterized by their $z$-invariant 
relative permittivity $\varepsilon(\B r)$ and relative permeability $\mu(\B r)$, where $\B r=(x,y)^{\rm T}$ is the position 
vector. This medium is illuminated by a monochromatic electromagnetic wave of 
pulsation $\omega=k_0/c$, amplitude $A_0(\B r,k_0)$ and phase $\phi_0(\B r,k_0)$ whose electric 
 field is linearly polarized along the $z$ axis, which is the so called Transverse Electric (TE) polarization,
 so that $\B E=E_z \B z$. 
Under these conditions, Maxwell's equations can be recast as the scalar wave equation: 
\begin{equation}
 \nabla\cdotp\left(\frac{1}{\mu}\,\nabla E_z\right) + k_0^2\,\varepsilon\, E_z = 0 ,
 \label{waveEq}
\end{equation}
By writing the total electric field in polar form as $E_z=A\expo^{\ic \phi}$ ($A$ and $\phi$ real), Eq.~(\ref{waveEq}) is 
separated into the following two equations:
  \begin{empheq}[left=\empheqlbrace]{align}
     &\B\nabla\cdotp\left(\frac{A^2}{\mu}\B\nabla \phi\right)=0    \label{eqsimon1} \\
  &( \B{\nabla}\phi)^2-k_0^2\varepsilon\mu-\frac{\B\nabla^2 A}{A}+ \frac{\B\nabla \mu}{\mu}\cdotp\frac{\B\nabla A}{A}=0   \label{eqsimon2}
  \end{empheq}
The physical meaning of these two equation is well known: 
the first is the continuity equation for the Poynting vector, 
while the second is the \emph{exact} eikonal equation governing the motion of the rays \cite{holland1995quantum,PhilbinExactGO}. 
They are usually solved through setting $\varepsilon$ and 
$\mu$ as known quantities and then solving for $E_z$, \textit{i.e.} $A$ and $\phi$. 
However, the methodology presented here allows us to fix arbitrarily two parameters and then compute 
the two others using Eqs. (\ref{eqsimon1})-(\ref{eqsimon2}).\\
From now on we consider an incident homogeneous plane wave with constant amplitude $A_0$ and 
phase $\phi_0(\B r,k_0)=k_0 \B n \cdotp \B r$, with $\B n =(\cos\theta_0,\sin\theta_0)^{\rm T}$ the unit vector defining the incidence direction. 
The gradient of the phase can then be written as 
$$\B{\nabla}\phi=\B n k_0+\B{\nabla}\psi,$$ where $\psi$ is an additional phase term. 
If $\B{\nabla}\psi\rightarrow 0$ and 
$A\rightarrow A_0$ as $r=\sqrt{x^2+y^2}\rightarrow +\infty$, 
the incident wave remains plane and the material will be invisible. \\

\section{Controlling amplitude and permeability}
In this section we suppose that we fix $A$ and $\mu$. 
Substituting $\B{\nabla}\phi$ into Eq.~(\ref{eqsimon1}), 
we obtain the following Poisson's equation for $\psi$
\begin{equation}
 \B\nabla\cdotp\left(\frac{A^2}{\mu}\B\nabla \psi\right)= -k_0\B n\cdotp\B\nabla\left(\frac{A^2}{\mu}\right) ,
 \label{poisson}
\end{equation}
which can be solved to give
$$ \B \nabla \psi(\B r)  =  -\frac{\mu(\B r)k_0}{2\pi A^2(\B r)} 
\int {\rm d}^2\B r'\frac{\B r-\B r'}{|\B r-\B r'|^2} \B n \cdotp 
\B{\nabla}'\left(\frac{A^2(\B r')}{\mu(\B r')}\right).$$
This shows that if we specify the quantity $\zeta=A^2/\mu$ over space then the gradient of the phase changes in response to the change 
in $\zeta$ in the same way the electric field responds to a charge density. 
Substituting the above equation into (\ref{eqsimon2}) then determines a relationship between $\varepsilon$ and $\mu$. \\
In the following we further assume that $A$ and $\mu$ are dispersionless and introduce the frequency 
independent quantities $\alpha=\phi/k_0$ and $\beta=\psi/k_0$. 
Locally, the permittivity dispersion takes the form of a lossless Drude model 
\begin{equation}
\varepsilon(\omega)=\varepsilon_{\infty}-\omega_{\rm p}^2/\omega^2,
\label{drudemodel}
\end{equation}
with the 
permittivity at infinite frequency $\varepsilon_{\infty}$ and the plasma frequency $\omega_{\rm p}$ defined as:
\begin{align}
\varepsilon_{\infty}&= \frac{(\B{\nabla} \alpha)^2}{\mu}=\frac{1}{\mu}\left[1+(\B{\nabla} \beta)^2+2\,\B n \cdotp \B{\nabla} \beta \right],\label{drudeparam1}\\
\omega_{\rm p}^2&=\frac{c^2}{\mu} \left(\frac{\B\nabla^2 A}{A}-\frac{\B\nabla \mu}{\mu}\cdotp\frac{\B\nabla A}{A}\right).
\label{drudeparam2}
\end{align}
The obtained permittivity is linear, spatially varying, with a $1/\omega^2$ dispersion and non-local since 
$\varepsilon_{\infty}$ depends on the incidence direction $\B n$. 
On the basis of time reversal, a plane wave coming from the opposite direction 
gives a total field with the same amplitude but an opposite phase as $\phi(-\B n)=-\phi(\B n)$, while invisibility 
is maintained for the same permittivity 
since $\varepsilon(-\B n)=\varepsilon(\B n)$, even if generally the amplitude and material profiles do not possess 
any particular symmetry.\\

\subsection{A special case}
There is a particular situation for which we can get rid of the non-locality, and this happens when 
$\B{\nabla}\beta=\B 0$, \textit{i.e.} when $\mu$ is proportional to $A^2$. In this case and 
in the ray optics approximation we retrieve a medium with 
unit index of refraction because $\varepsilon\rightarrow 1/\mu$ as $\omega\rightarrow +\infty$, 
which is an inhomogeneous medium where all the waves travel in straight lines and without reflection. 
Essentially, our approach can be understood by considering this limiting case $\varepsilon=1/\mu$ and extending 
it to work for all frequencies and all incidences by adding dispersive and non-local terms into $\varepsilon$.
On the other side of the spectrum, 
the medium becomes singular in the quasi-static limit since $\abs{\varepsilon}\rightarrow +\infty$ as $\omega\rightarrow 0$. 
This behaviour is due to the fact that any permeability inhomogeneity will cause 
large scattering at low frequencies, and one needs large changes in the permittivity to counteract this.\\
Without loss of generality, we now consider the case where $\mu=A^2$: this 
implies that the phase is exactly given by $\B{\nabla}\phi=\B n k_0$ everywhere, 
\textit{i.e.} the field is a plane wave with a non-uniform amplitude, and the Drude parameters simplify as
\begin{equation}
\varepsilon_{\infty}=\frac{1}{\mu}\quad \text{and} \quad
\omega_{\rm p}^2 =\frac{c^2}{\mu} \left( \frac{\B\nabla^2\sqrt{\mu}}{\sqrt{\mu}}-\frac{\B\nabla \mu}{\mu}\cdotp\frac{\B\nabla \sqrt{\mu}}{\sqrt{\mu}}\right). \label{equchi}
\end{equation}
We note that in this case, $\varepsilon$ is frequency dispersive but does not depend on the incidence angle, 
similarly to the P\"{o}schl-Teller profile 
(which is reflectionless for all angles and depends on $\omega$, see e.g. \cite{Lekner2007}) as 
the permittivity is analogous to the quantum potential for the Shr\"{o}dinger equation.\\
\begin{figure}[t]
 \centering
  \includegraphics[width=1\columnwidth]{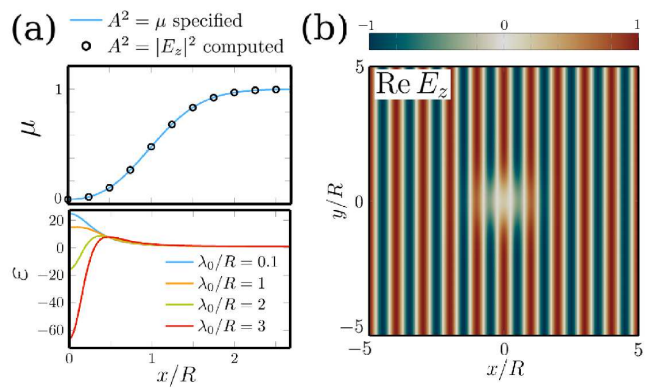}
 \caption{Invisible material in the case $\mu=A^2$ with $80\%$ damping of the field in the centre. 
 (a) Permeability (top) and permittivity (bottom) profiles along the radial direction. 
(b) Real part of the electric field $E_z$ for $\lambda_0/R=1$. \label{fig1}}
\end{figure}
As an example, suppose we want to obtain a field with a prescribed Gaussian amplitude $A=1-f\exp(-r^2/R^2)$,
and that $\mu=A^2$ (see blue line on the top panel of Fig.~\ref{fig1}~(a)), with 
$R=700\,$nm and $f=0.8$. Note that this results in a permeability profile with values below unity, which seems to contradicts our 
assumption of neglecting frequency dispersion for $\mu$. In practice indeed we would likely only be able to realise 
the $\mu$ profile containing regions of $\mu<1$ for one single frequency. 
The calculated permittivity profile is shown for several wavelengths on Fig.~\ref{fig1}~(a) (bottom panel). As discussed previously, 
the required $\varepsilon$ is roughly equal to $1/\mu$ for $\lambda_0/R=0.1$, while one needs more extreme permittivity values at longer wavelengths. 
We solved the wave equation (\ref{waveEq}) using a Finite Element Method (FEM) for $\lambda_0/R=1$, with a plane wave of unit amplitude incident from the negative $x$ axis 
and Perfectly Matched Layers (PML) to truncate the domain. 
The real part of the electric field $E_z$ is plotted on Fig.~\ref{fig1}~(b) and reveals a clear damping of the field as well as no scattering and a planar 
wavefront everywhere. The computed square norm of the field matches the required one perfectly 
(see black circles on the top panel of Fig.~\ref{fig1}~(a)).\\
Note that the Transverse Magnetic (TM) polarization case can be treated similarly by replacing $E_z$ by $H_z$ and swapping $\varepsilon$ and $\mu$.

\subsection{The non-magnetic case}
\begin{figure}[t]
 \centering
 \includegraphics[width=1\columnwidth]{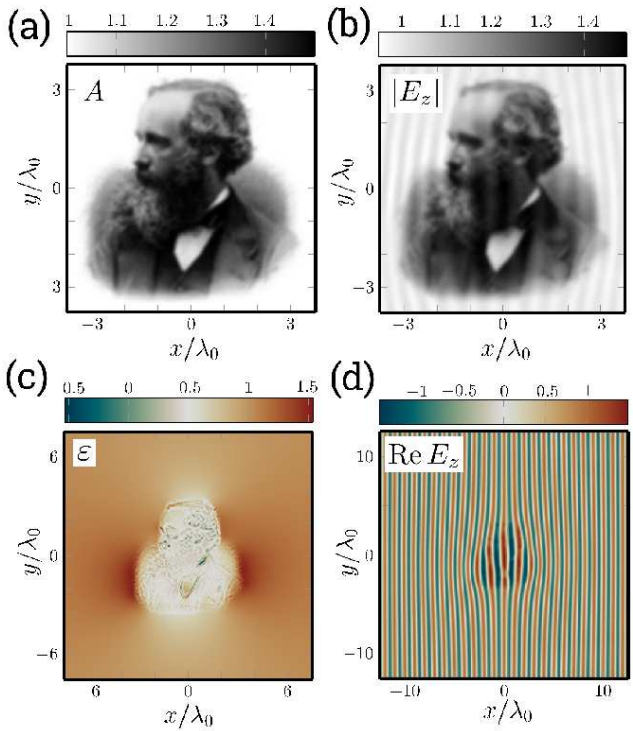}
 \caption{Invisible material profile in the non-magnetic case ($\mu=1$) with arbitrary control of the amplitude. 
(a) Specified amplitude (from a picture of James Clerk Maxwell). 
(b) Computed amplitude. 
(c) Permittivity profile. 
(d) Real part of the electric field, showing the invisibility effect. 
\label{fig2}}
\end{figure}
For practical reasons, we investigate the possibility 
of having non-magnetic invisible profiles ($\mu=1$). 
We solve Eq.(\ref{eqsimon1}) to obtain the phase and the parameters for the permittivity reduce to:
\begin{equation}
\varepsilon_{\infty}= 1+(\B{\nabla} \beta)^2+2\,\B n \cdotp \B{\nabla} \beta \quad \text{and} \quad
\omega_{\rm p}^2= c^2 \frac{\B\nabla^2 A}{A}
\label{epsi_diel_Eq}
\end{equation}
\begin{figure}[t]
 \centering
 \includegraphics[width=0.85\columnwidth]{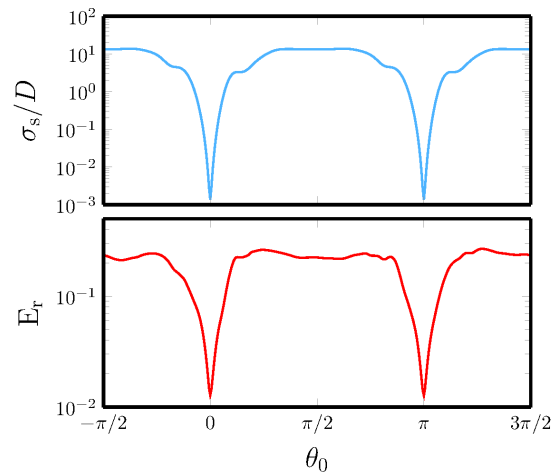}
 \caption{Angular response of the permittivity profile of Fig.~\ref{fig2}~(c). Top: scattering cross section $\sigma_{\rm s}$ 
 normalized to the profile size $D$. Bottom: average error on the amplitude ${\rm E_ r}$ defined by Eq.~(\ref{error_Eq}).
\label{fig2bis}}
\end{figure}
\begin{figure*}[t]
 \centering
 \includegraphics[width=1\textwidth]{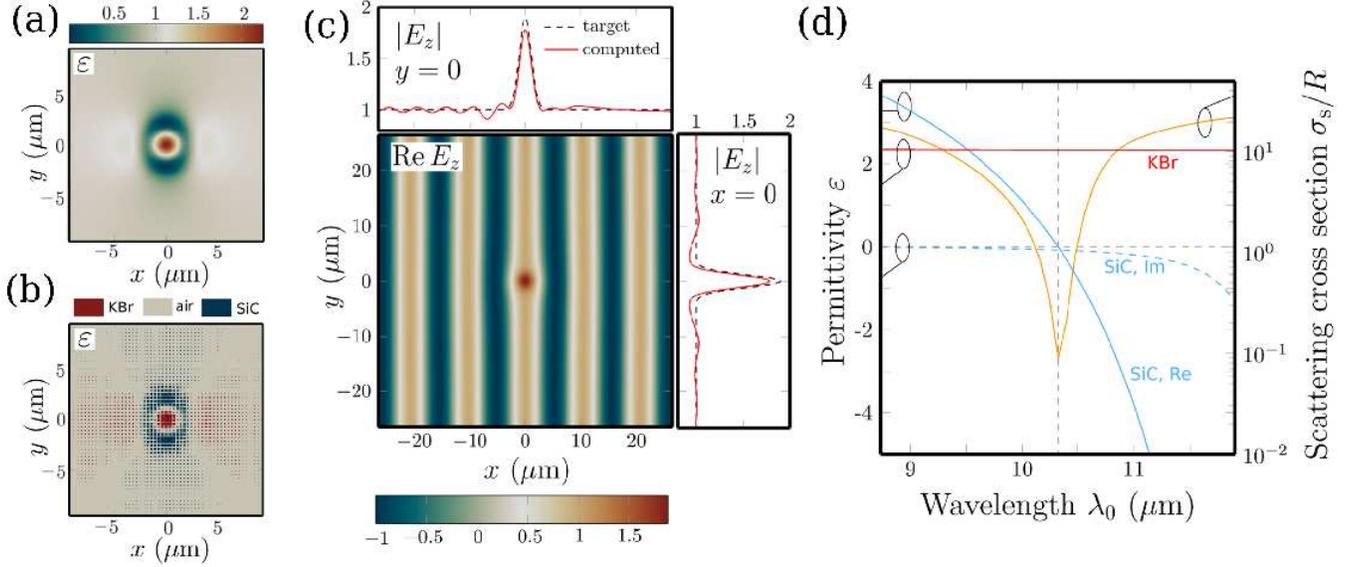}
 \caption{Invisible metamaterial with sub-wavelength control of the amplitude. (a) Continuous and (b) metamaterial permittivity 
 profiles. (c) Central colour map: real part of the electric field at $\lambda_0=\SI{10.32}{\micro\meter}$, top and right panels: 
 target (black dashed lines) and calculated (red solid lines) amplitudes for $y=0$ and $x=0$ respectively. 
 (d) Left ordinate axis: permittivity dispersion of SiC (solid and dashed cyan lines for real and imaginary parts) and KBr (solid red line),
 the horizontal dashed line indicates a zero value; 
 right ordinate axis: scattering cross section spectra of the metamaterial structure. 
 The vertical dashed line indicates $\lambda_0=\lambda_{\rm L}=\SI{10.32}{\micro\meter}$ at which we designed the 
 structure.
\label{fig3}}
\end{figure*}
To illustrate the arbitrariness of the choice of the amplitude, we used a profile extracted 
from a grayscale image of James Clerk Maxwell depicted on Fig.~\ref{fig2}~(a), where dark values correspond
to a 50\% enhancement of the field, with a lateral ``size'' of approximately $D=6\lambda_0$. 
The permittivity profile is displayed on Fig.~\ref{fig2}~(c), and presents small features and rapidly varying values 
between $-0.5$ and $1.5$. 
The real part of $E_z$ is displayed on Fig.~\ref{fig2}~(d), and proves clearly that the field is not a plane wave, 
with a retarded phase on the left and an advanced phase on the right of the inhomogeneity, 
but that this profile does not induce any scattering. 
The required field enhancement is respected as can be seen on 
Fig.~\ref{fig2}~(b) with no more than 5\% relative error, 
albeit some small reflections due to numerical inaccuracies. This 
proves the ability of the method to devise invisible non-magnetic media capable of shaping intricate magnitude patterns. 
We then investigate the angular response of this permittivity profile in terms of invisibility an amplitude control. 
To quantify this, we computed the scattering cross section $\sigma_{\rm s}$ normalized to the profile size $D$, along 
with the average error on the amplitude ${\rm E_ r}$ defined as 
\begin{equation}
{\rm E_ r}(\theta_0)=\frac{1}{S_\Omega}\int_\Omega {\rm d} \B r\left\|1-\frac{|E_z(\theta_0)|}{A}\right\|
\label{error_Eq}
\end{equation}
where $\Omega=[24\lambda_0\times 24\lambda_0]$ is the computational window used (cf. Fig.~\ref{fig2}~(d)) with surface 
$S_\Omega = (24\lambda_0)^2$. The results are plotted as a function of the incident angle $\theta_0$ on Fig.~\ref{fig2bis}, 
and clearly indicate a strong reduction of the scattering and an accurate reconstruction of the field magnitude for 
the reference configuration ($\theta_0 = \pi$) as well as for the anti-parallel direction of incidence ($\theta_0 = 0$), as discussed before.
As expected, both effects are fairly narrow-band due to the non-locality of the permittivity.

\begin{figure*}[t]
 \centering
 \includegraphics[width=1\textwidth]{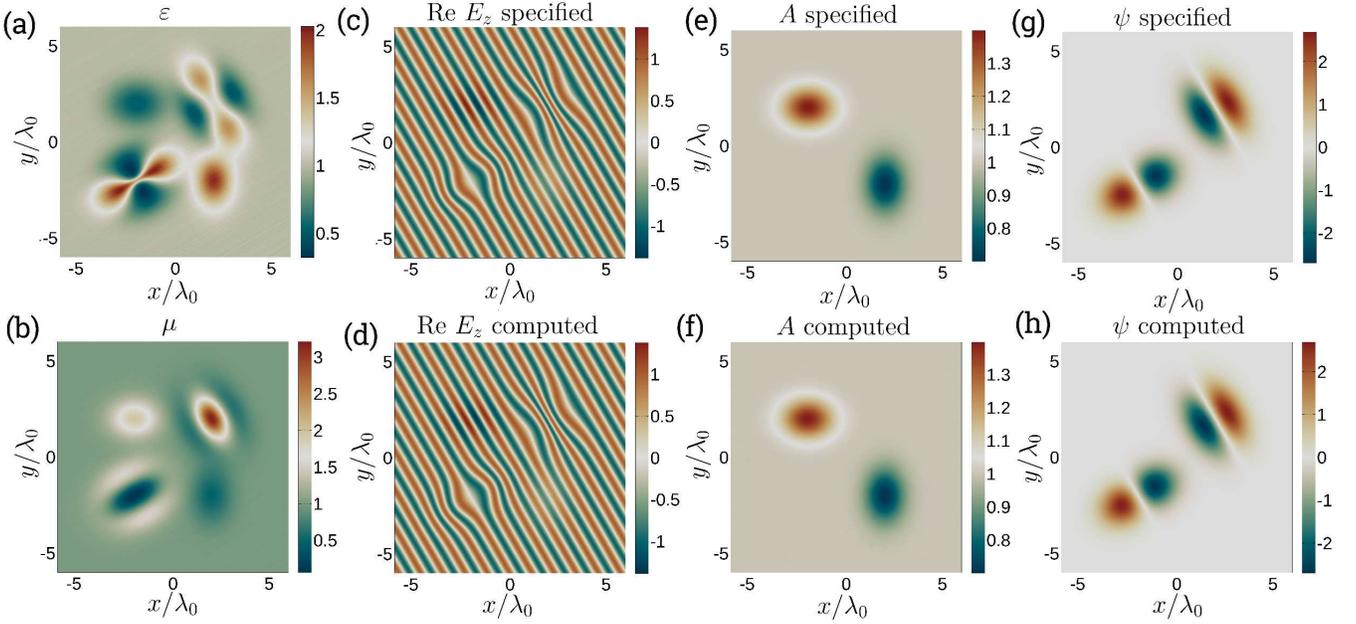}
 \caption{Inverse design of amplitude and phase profiles (see text for definitions) represented in (e), (g), giving a desired 
 electric field (c). Required permittivity (a) and permeability (b) are the used to solve the wave equation (direct problem) 
 for a sanity check of the field (d), amplitude (f) and phase (h).
\label{fig4}}
\end{figure*}

\subsection{Metamaterial implementation}
As for a possible experimental verification of our method, we propose a metamaterial structure that approximates the permittivity profile given by 
Eq.~(\ref{epsi_diel_Eq}) at $\lambda_0=\SI{10.32}{\micro\meter}$ with $A=1-f\exp(-r^2/R^2)$, $f=-0.9$, 
and $R=\lambda_0/6.5=\SI{1587}{\nano\meter}$. The resulting continuous permittivity profile is given on 
Fig.~\ref{fig3}~(a) and is varies between $0.044$ and $2.239$. To be able to reach values of permittivity smaller than 
unity, we use silicon carbide (SiC), a polaritonic material that has a strong dispersion 
in the thermal infrared range given by the Drude-Lorentz model \cite{palik} 
$\varepsilon_{\rm SiC}(\omega)=\varepsilon_{\infty}[1+(\omega_L^2-\omega_T^2)/(\omega_T^2-\omega^2+\ic\Gamma\omega)]$, 
with $\varepsilon_{\infty}=6.7$, $\omega_{\rm L}=\SI{1.82e14}{\radian\per\second}$, 
$\omega_{\rm T}=\SI{1.49e14}{\radian\per\second}$ and $\Gamma=\SI{8.96e11}{\radian\per\second}$ (see solid and dashed 
cyan lines on Fig.~\ref{fig3}~(d)). This material exhibits a dielectric to metallic transition 
around $\lambda_0=\lambda_{\rm L}=\SI{10.32}{\micro\meter}$ so that $\varepsilon_{\rm i}(\lambda_0)=0.0009 - 0.0815\ic$. 
For values greater than unity, we use potassium bromide (KBr) with permittivity $\varepsilon_{\rm KBr}(\lambda_0)=2.3280$ \cite{li1976refractive}. 
The hybrid metamaterial structure is a $51\times 51$ array of square unit cells of period $d=\lambda_0/27=\SI{377}{\nano\meter}$. 
The continuous map of Fig.~\ref{fig3}~(a) is discretized at the centre $(x_i,y_j)$ of those unit cells resulting in a discrete set of 
values $\varepsilon_{ij}=\varepsilon(x_i,y_j)$. 
Since the period is much smaller than the wavelength, we can safely use an effective permittivity $\varepsilon _{\mathrm {eff} }$ given by
the Maxwell-Garnett homogenization formula: 
$${\left({\frac {\varepsilon _{\mathrm {eff} }-\varepsilon _{\rm h}}{\varepsilon _{\mathrm {eff} }+2\,\varepsilon _{\rm h}}}\right)
=f\left({\frac {\varepsilon _{\rm i}-\varepsilon _{\rm h}}{\varepsilon _{\rm i}+2\,\varepsilon _{\rm h}}}\right)
} $$
where $\varepsilon _{\rm h}$ is the permittivity of the host medium (air in our case), 
$\varepsilon _{\rm i}$ is the permittivity of the inclusions (either SiC or KBr),
$f=a^2/d^2$ is the filling fraction and $a$ is the length of the square section of the rods. The structure is then 
constructed as follows: if $\varepsilon_{ij}<0.99$ we use SiC rods, if $\varepsilon_{ij}>1.01$ we use KBr rods, otherwise we just use air 
(see Fig.~\ref{fig3}~(b)). The real part of the electric field is plotted on Fig.~\ref{fig3}~(c), and clearly illustrates the 
invisibility effect and the sub-wavelength control of the amplitude. The top and left panels compare the target (black dashed lines) 
and calculated (red solid lines) amplitudes for $y=0$ and $x=0$ respectively, 
revealing a quasi perfect match apart from a small scattering, mostly due to the truncation and discretization of the permittivity profile and 
a slightly weaker amplitude than expected, due to losses in SiC rods. The scattering cross section spectrum on Fig.~\ref{fig3}~(d) exhibits 
a pronounced dip around $\lambda_0=\SI{10.32}{\micro\meter}$, which illustrates the strong reduction of diffraction resulting in a quasi-invisible 
complex metamaterial.\\

\section{The inverse problem: controlling amplitude and phase}
Finally, we study the inverse problem of finding invisible material properties that give a pre-defined electric field. 
To this aim, we fix the amplitude $A$ and the additional phase term $\psi$ and rewrite Eq.~(\ref{eqsimon1}) as:
\begin{equation}
A^2\B\nabla \phi \cdotp\B\nabla u =  \B\nabla\cdotp\left(A^2\B\nabla \phi\right),
\end{equation}
with $u=\ln\mu$. This equation is then solved numerically and the obtained value of $\mu$ is plugged into Eq.~(\ref{eqsimon2}) 
to obtain $\varepsilon$.\\
For the following example, we set $\lambda_0=\SI{700}{\nano\meter}$, $R=\lambda_0$, $\theta_0=\pi/3$, 
\begin{align*}
A=1 & - 0.3\,\expo^{-\left[(x-2\lambda_0)^2+0.5(y+2\lambda_0)^2\right]/R^2} \\
& + 0.4\,\expo^{-\left[0.6(x+2\lambda_0)^2+(y-2\lambda_0)^2\right]/R^2}
\end{align*}
and 
$$
\psi = k_0\left[x''_a\,\expo^{-\left[{x''_a}^2+0.4{y''_a}^2\right]/R^2} - 0.7\,x''_b\,\expo^{-\left[0.5{x''_b}^2+{y''_b}^2\right]/R^2}\right]
$$
using the shifted and rotated coordinates:
\begin{align*}
&x''_a = n_x x'_a + n_y y'_a, \qquad &x'_a= x-2\lambda_0, \\
&y''_a = -n_y x'_a + n_x y'_a,\qquad &y'_a = y-2\lambda_0,\\
&x''_b = n_x x'_b + n_y y'_b ,\qquad  &x'_b= x+2\lambda_0,\\
&y''_b = -n_y x'_b + n_x y'_b ,\qquad &y'_b = y+2\lambda_0 .
\end{align*}
This particular choice of amplitude and phase will give the following wave behaviour:
amplitude damping at $(+2\lambda_0,-2\lambda_0)$,
amplitude enhancement at $(-2\lambda_0,+2\lambda_0)$,
phase expansion at $(-2\lambda_0,-2\lambda_0)$ and 
phase compression at $(+2\lambda_0,+2\lambda_0)$ (see Figures~\ref{fig4} (e), (g) and (c) for the specified amplitude, 
additional phase and electric field respectively). The obtained value of material properties are 
plotted on Figs.~\ref{fig4} (a) for the permittivity and (b) for the permeability. 
These non trivial profiles allow us to control the wave propagation quite arbitrarily in the near field 
while being transparent to a specific incident plane wave. Note that as stated before, the same profiles are still invisible
for a wave coming from the opposite direction, and maintain the amplitude control but the phase has now opposite sign.\\
To double check the validity of our results, we solved the wave equation (\ref{waveEq}) employing the permittivity and permeability 
obtained by our approach. The results are plotted in Figs.~\ref{fig4} (f), (h) and (d) for the amplitude, 
additional phase and electric field respectively and match the required wave behaviour perfectly. The generality of this 
inverse problem makes it quite versatile and reveals a family of amplitude and phase controlling invisible electromagnetic media.

\section*{Conclusion}
In conclusion, we have presented a flexible and systematic methodology to derive isotropic and lossless material properties needed to manipulate 
the amplitude and phase of the electromagnetic field in an arbitrary way, for planar propagation. 
In addition, our work provides a contribution in the understanding of what 
governs scattering in this type of media. 
Since it is based on the scalar wave equation, it could be easily extended to other fields such acoustics or fluid dynamics. 
In particular we have applied this method to derive a large class of invisible permittivity and permeability profiles. 
We illustrated these concepts through numerical examples for TE polarized plane waves 
using both $\varepsilon$ and $\mu$ and obtained omni-directional 
invisibility and control of the amplitude. Then we studied the case of non-magnetic materials and showed that one can obtain invisibility 
and fashion the spatial variation of the magnitude of the electric field for two anti-parallel directions of incidence. 
A metamaterial structure working in the infrared has been proposed, exhibiting sub-wavelength control of waves and invisibility at the 
same time. Finally, we tackled the inverse problem of finding non-scattering material properties that give a specified electric 
field. These results pave the way for a new route towards achieving invisibility with isotropic materials, 
and may offer an alternative paradigm for the design of nanophotonic devices with enhanced performances.\\

\begin{acknowledgments}
This work was funded by the Engineering and Physical
Sciences Research Council (EPSRC), UK, under a
Programme Grant (EP/I034548/1) ``The Quest for
Ultimate Electromagnetics using Spatial Transformations
(QUEST)''.
\end{acknowledgments}


%

\end{document}